\documentclass[twoside]{dis08}
\usepackage[latin1]{inputenc}
\usepackage[dvips]{graphicx,epsfig,color}
\usepackage{wrapfig,rotating}
\usepackage{amssymb,amsmath,array}

\begin{document}
\title{Extraction of the proton parton density functions using a NLO-QCD fit of
the combined H1 and ZEUS inclusive DIS cross sections}

\author{A M Cooper-Sarkar on behalf of the ZEUS and H1 collaborations
%
%
\vspace{.3cm}\\
%
Oxford University - Dept of Physics \\
Denys Wilkinson Bdg, Oxford, OX1 3RH - UK
%
}

\maketitle

\begin{abstract}
The combined HERA-I data set, of neutral and charged current 
inclusive cross-sections for $e^+p$ and $e^-p$ scattering, is used as the
sole input for a next-to-leading order (NLO) QCD parton distribution 
function (PDF) fit. 
The consistent treatment of systematic uncertainties in the joint data set 
ensures that experimental uncertainties on the PDFs can be calculated
without 
need for an increased $\chi^2$ tolerance. This results in PDFs with greatly
reduced experimental uncertainties compared to the separate analyses of 
the ZEUS and H1 experiments. Model uncertainties, including those 
arising from parametrization dependence, are also carefully considered. The
resulting HERAPDFs have impressive precision compared to the global fits.
\end{abstract}

\section{Introduction}

\label{sec:intro}
The kinematics
of lepton hadron scattering is described in terms of the variables $Q^2$, the
invariant mass of the exchanged vector boson, Bjorken $x$, the fraction
of the momentum of the incoming nucleon taken by the struck quark (in the 
quark-parton model), and $y$ which measures the energy transfer between the
lepton and hadron systems.
The differential cross-section for the neutral current (NC) process is given in 
terms of the structure functions by
\[
\frac {d^2\sigma(e^{\pm}p) } {dxdQ^2} =  \frac {2\pi\alpha^2} {Q^4 x}
\left[Y_+\,F_2(x,Q^2) - y^2 \,F_L(x,Q^2)
\mp Y_-\, xF_3(x,Q^2) \right],
\]
where $\displaystyle Y_\pm=1\pm(1-y)^2$. 
The structure functions $F_2$ and $xF_3$ are 
directly related to quark distributions, and their
$Q^2$ dependence, or scaling violation, 
is predicted by perturbative QCD. For low $x$, $x \leq 10^{-2}$, $F_2$ 
is sea quark dominated, but its $Q^2$ evolution is controlled by
the gluon contribution, such that HERA data provide 
crucial information on low-$x$ sea-quark and gluon distributions.
At high $Q^2$, the structure function $xF_3$ becomes increasingly 
important, and gives information on valence quark distributions. 
The charged current (CC) interactions also
enable us to separate the flavour of the valence distributions 
at high-$x$, since their (LO) cross-sections are given by, 
\[
\frac {d^2\sigma(e^+ p) } {dxdQ^2} = \frac {G_F^2 M_W^4} {(Q^2 +M_W^2)^2 2\pi x}
x\left[(\bar{u}+\bar{c}) + (1 - y)^2 (d + s) \right],
\]
\[
\frac {d^2\sigma(e^- p) } {dxdQ^2} = \frac {G_F^2 M_W^4} {(Q^2 +M_W^2)^2 2\pi x}
x\left[(u + c) + (1 - y)^2 (\bar{d} + \bar{s}) \right].
\]

Parton Density Function (PDF) determinations are usually obtained from global NLO 
QCD fits~\cite{mrst,cteq,zeus-s}, which use fixed target 
DIS data as well as HERA data. In such analyses, the high statistics HERA NC 
$e^+p$ data have determined the low-$x$ sea and 
gluon distributions, whereas the fixed target data have determined 
the valence distributions. Now that high-$Q^2$ HERA data on NC and CC
 $e^+p$ and $e^-p$ inclusive double 
differential cross-sections are available, PDF fits can be made to HERA 
data alone, since the HERA high $Q^2$ cross-section 
data can be used to determine the valence distributions. This has the 
advantage that it eliminates the need for heavy target corrections, which 
must be applied to the $\nu$-Fe and $\mu D$ fixed target data. Furthermore
there is no need to assume isospin symmetry, i.e. that $d$ in the 
proton is the same as $u$ in the neutron, 
since the $d$ distribution can be obtained directly from CC $e^+p$ data. 

The H1 and ZEUS collaborations have both used their data to make PDF 
fits~\cite{zeusj},~\cite{h1}. Both of these data sets
have very small statistical uncertainties, so that the contribution of 
systematic uncertainties becomes dominant and consideration of 
point to point correlations between systematic uncertainties is essential.
The ZEUS analysis takes account of correlated experimental 
systematic errors by the Offset Method, whereas H1 uses the Hessian 
method~\cite{durham}. 
Whereas the resulting ZEUS and H1 PDFs are compatible, the gluon PDFs do 
have rather different shapes, see Fig.~\ref{fig:summary}, and the 
uncertainty bands spanned by these analyses 
are comparable to those of the global fits.

It is possible to improve on this situation 
since ZEUS and H1 are measuring the same physics in the same 
kinematic region. These data have been combined them using a 'theory-free' 
Hessian fit in which the only assumption is that there is a true 
value of the cross-section, for each process, at each $x,Q^2$ 
point~\cite{combination},~\cite{Feltesse}. 
The resulting systematic uncertainties 
on each of the combined data points are significantly smaller than the 
statistical errors. In the present paper this combined data set is used as the 
input to a NLO QCD PDF fit. The consistency of the input data set 
and its small systematic uncertainties enable us 
to calculate the experimental uncertainties on the PDFs using the 
$\chi^2$ tolerance, $\Delta\chi^2=1$. This represents a further advantage 
compared to those global fit analyses where increased tolerances of 
$\Delta\chi^2=50-100$ are used to account for data inconsistencies. 

For the present HERAPDF0.1 fit, the role of correlated 
systematic uncertainties is no longer crucial since these uncertainties are 
relatively small. This ensures that similar results are 
obtained using either Offset or Hessian methods, or by simply combining 
statistical and systematic uncertainties in quadrature. 
For our central fit we have 
chosen to combine the 43 systematic uncertainties 
which result from the separate ZEUS and H1 data sets in quadrature, 
and to Offset the 4 sources of uncertainty which result from the combination 
procedure. This results in the most conservative uncertainty estimates on the 
resulting PDFs. 

Despite our conservative procedure
 the experimental uncertainties on the resulting 
PDFs are impressively small and a thorough consideration of further 
uncertainties 
due to model assumptions is necessary. In section~\ref{sec:anal} we describe 
the NLO QCD analysis and model assumptions. In section~\ref{sec:results} 
we give results and in section~\ref{sec:conc} we give a summary.

\section{Analysis}
\label{sec:anal}
 
The QCD predictions for the structure functions 
are obtained by solving the DGLAP evolution equations 
at NLO in the MSbar scheme with the
renormalisation and factorization scales chosen to be $Q^2$.
The DGLAP equations yield the PDFs
 at all values of $Q^2$ provided they
are input as functions of $x$ at some input scale $Q^2_0$. This scale has been 
chosen to be $Q^2_0 = 4$GeV$^2$ and variation of this choice is considered 
as one of the model uncertainties.
The resulting PDFs are then convoluted with NLO coefficient functions to 
give the structure functions which enter into the expressions for the 
cross-sections. The choice of the heavy quark 
masses is, $m_c=1.4, m_b=4.75$GeV, and variation of these 
choices is included in 
the model uncertainties. For this preliminary analysis, the heavy quark 
coefficient functions have been caluclated in the zero-mass variable flavour 
number scheme. The strong coupling constant was fixed to 
$\alpha_s(M_Z) =  0.1176$~\cite{PDG}, and variations in this value of $\pm 0.002$ 
have also been considered.

The fit is made at leading twist. 
The HERA data have of the invariant mass of the hadronic system,$W^2$, 
of $W^2_{min} = 300$GeV$^2$ and maximum $x$, $x_{max} = 0.65$, 
such that they are in a kinematic region where there is no
sensitivity to target mass and large-$x$ higher 
twist contributions. However a minimum $Q^2$ cut is imposed 
to remain in the kinematic region where
perturbative QCD should be applicable. This has been chosen such that
 $Q^2_{min} = 3.5$~GeV$^2$. Variation of this cut is included as one
 of the model uncertainties. 

 A further 
model uncertainty is the choice of the initial parametrization at 
$Q^2_0$. The PDFs are parametrized by the generic form 
\begin{equation}
 xf(x) = A x^{B} (1-x)^{C} (1 + D x + E x^2 +F x^3),
\label{eqn:pdf}
\end{equation}
and the number of parameters is chosen by 'saturation of the $\chi^2$', 
such that parameters $D,E,F$ are only varied if this brings significant 
improvement to the $\chi^2$. Otherwise they are set to zero.

For our central fit,
the PDFs which are parametrized are $xu_v$, $xd_v$, $xg$ and 
$x\bar{U}$, $x\bar{D}$. 
The normalisation parameters, $A$, for the $d$ and $u$ valence are 
constrained to impose the number sum-rules and the normalisation parameter $A$
for the gluon is constrained to impose the momentum sum-rule. 
The $B$ parameters which constrain the low-$x$ behaviour of the $u$ and $d$ 
valence distributions are set equal, and  
the $B$ parameters are also set equal for $x\bar{U}$ and $x\bar{D}$, such that 
there is a single $B$ parameter for the valence and another different single 
$B$ parameter for the sea distributions. 
Assuming that the strange and charm quark distributions can be expressed as 
$x$ independent fractions, $f_s=0.33$ and $f_c=0.15$, of the $d$ and $u$ 
type sea, 
gives the further constraint $A(\bar{U})=A(\bar{D}) (1-f_s)/(1-f_c)$. 
The value of $f_s=0.33$ has been chosen to be consistent with determinations 
of this fraction using neutrino induced di-muon production. This value
 has been varied to evaluate model
uncertainties. The charm fraction has been set to be consistent with dynamic
 generation of charm
from the start point of $Q^2= m_c^2$ in a zero-mass-variable-flavour-number 
scheme. A small variation of the value of 
$f_c$ is included in the model uncertainties.
 Saturation of the $\chi^2$ leads us to set the 
parameters $D,E,F=0$, for all partons except 
$xu_v$ for which only $F=0$.

The results are presented using this parametrization, including six sources 
of model uncertainty due to variation of: $m_c,m_b,f_s,f_c,Q^2_0, Q^2_{min}$.
Comparison is made to three other classes of parametrization, one based on the
ZEUS-JETs parametrization~\cite{zeusj}, one based on the H1 
parametrization~\cite{h1} and one based on the current parametrization but 
allowing $D \not= 0$ for the gluon. Comparison is also made 
to results obtained by varying 
$\alpha_s(M_Z)$,  see reference~\cite{url} for details. 
Our central choice has less model 
dependence than the ZEUS-Style parametrization because it has fewer asumptions 
concerning $\bar d- \bar u$, and it has less model dependence than the
 H1-style 
parametrization in that it does not assume equality of all $B$ parameters.
Furthermore, although all types of parametrization give acceptable $\chi^2$ 
values, the central parametrization has the best $\chi^2$ and 
it gives the most conservative experimental errors.

\section{Results}
\label{sec:results}

\begin{figure}[tbp]
\vspace{-0.2cm} 
\centerline{
\epsfig{figure=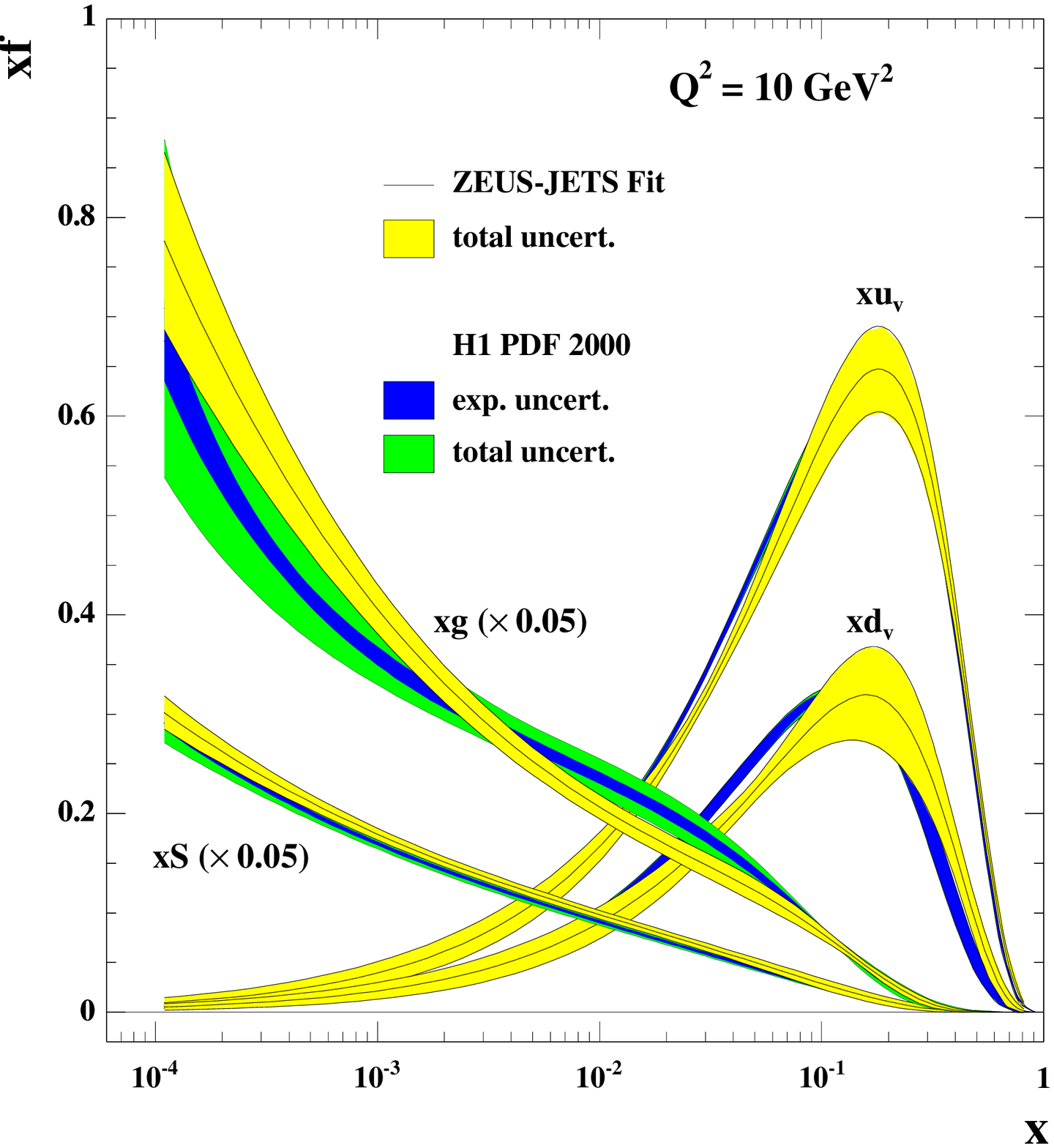,width=0.45\textwidth,height=6.1cm}
\epsfig{figure=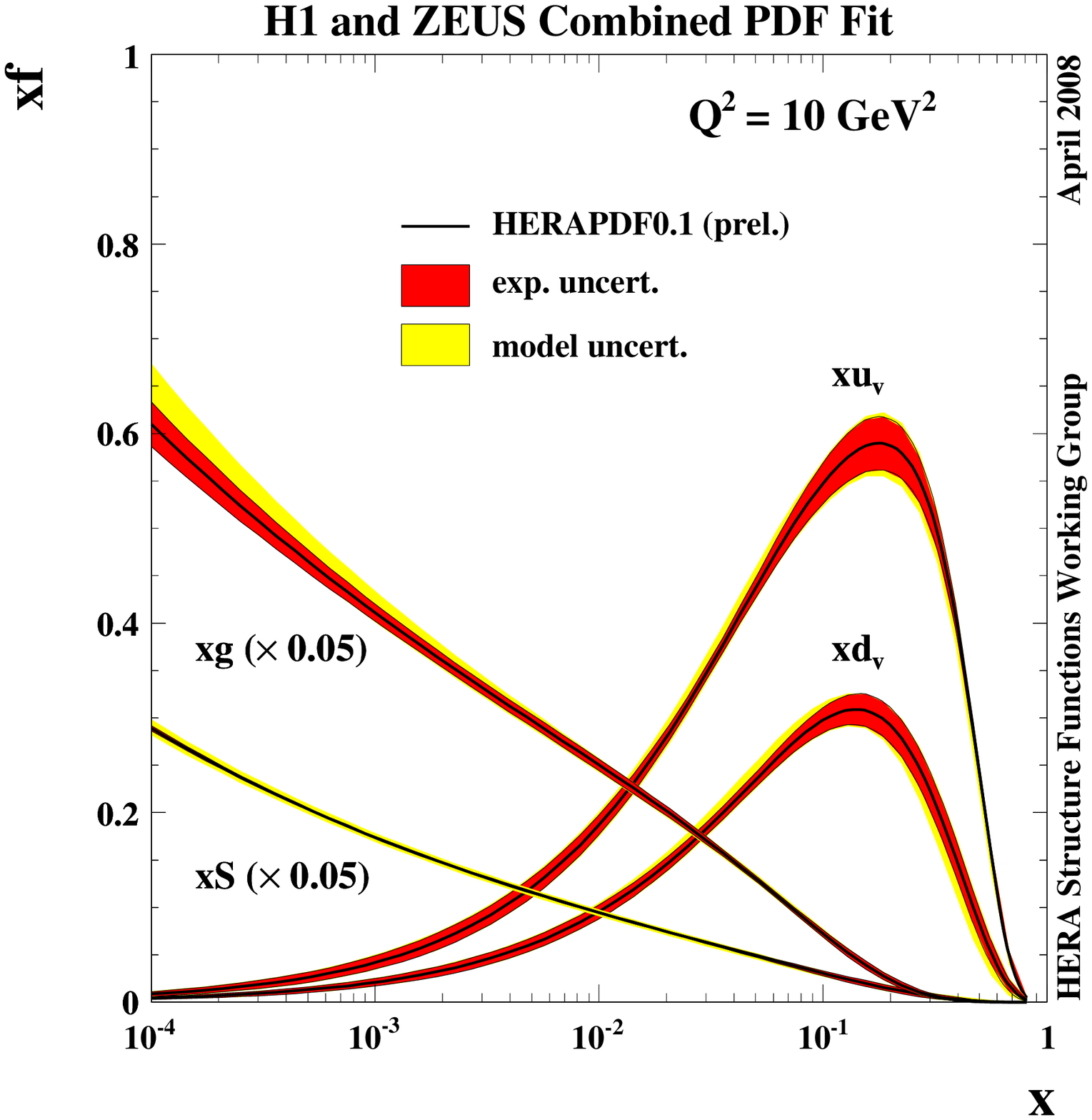 ,width=0.5\textwidth,height=6.5cm}}
\caption {Left: PDFs from the ZEUS-JETS and H1PDF2000 PDF separate 
analyses of ZEUS and H1. Right: HERAPDF0.1 PDFs
 from the analysis of the combined data set}
\label{fig:summary}
\end{figure}

The total uncertainty of the PDFs obtained from the HERA combined data set is 
much reduced compared to the PDFs extracted from the analyses of the separate 
H1 and ZEUS data sets, as can be seen from the 
summary plots in Fig.~\ref{fig:summary}, where these new HERAPDF0.1 
PDFs are compared to the ZEUS-JETS and H1PDF2000 PDFs.
  
\begin{figure}[tbp]
\vspace{-0.5cm} 
\centerline{
\epsfig{figure=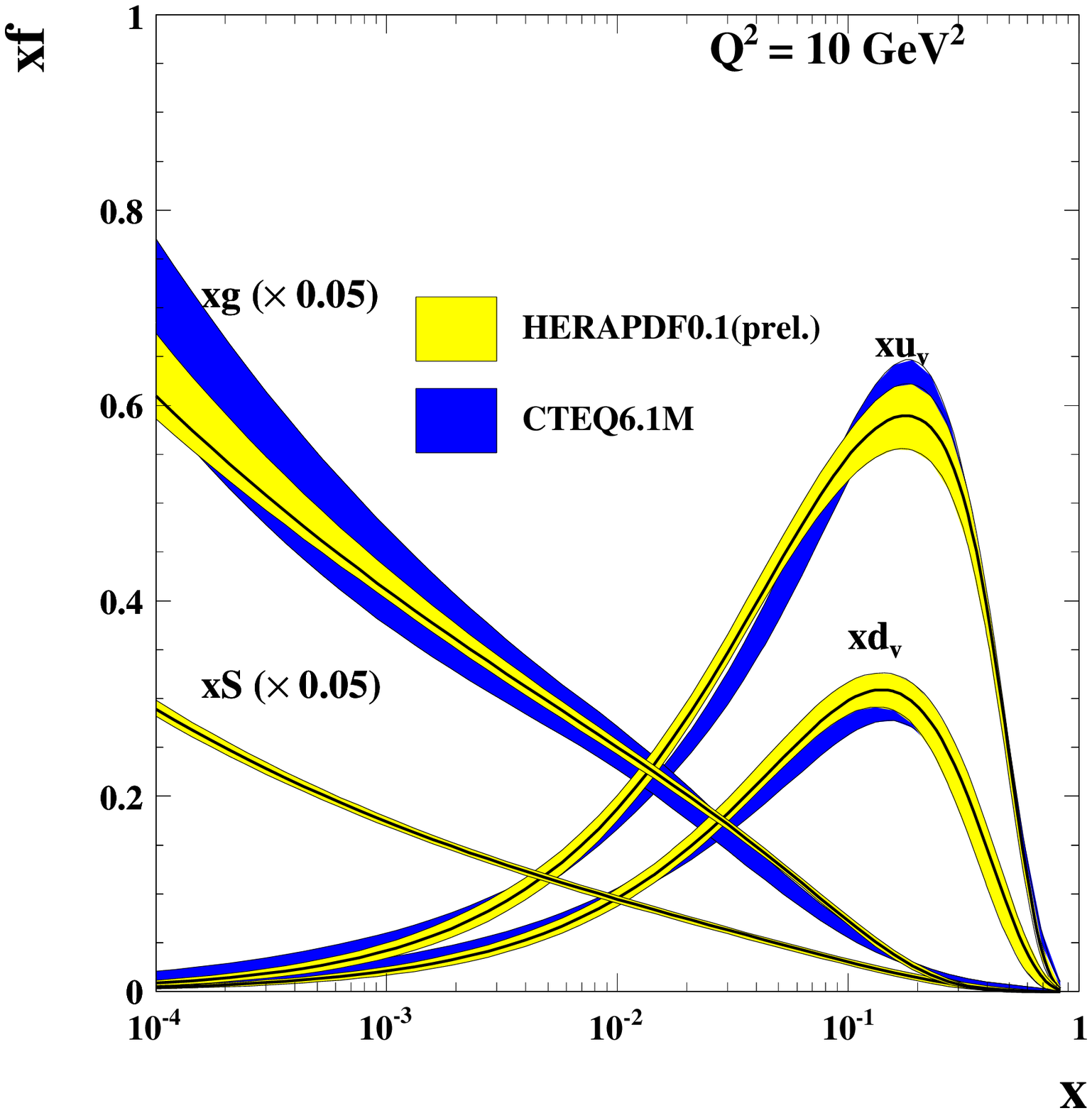 ,width=0.5\textwidth,height=6.5cm}
\epsfig{figure=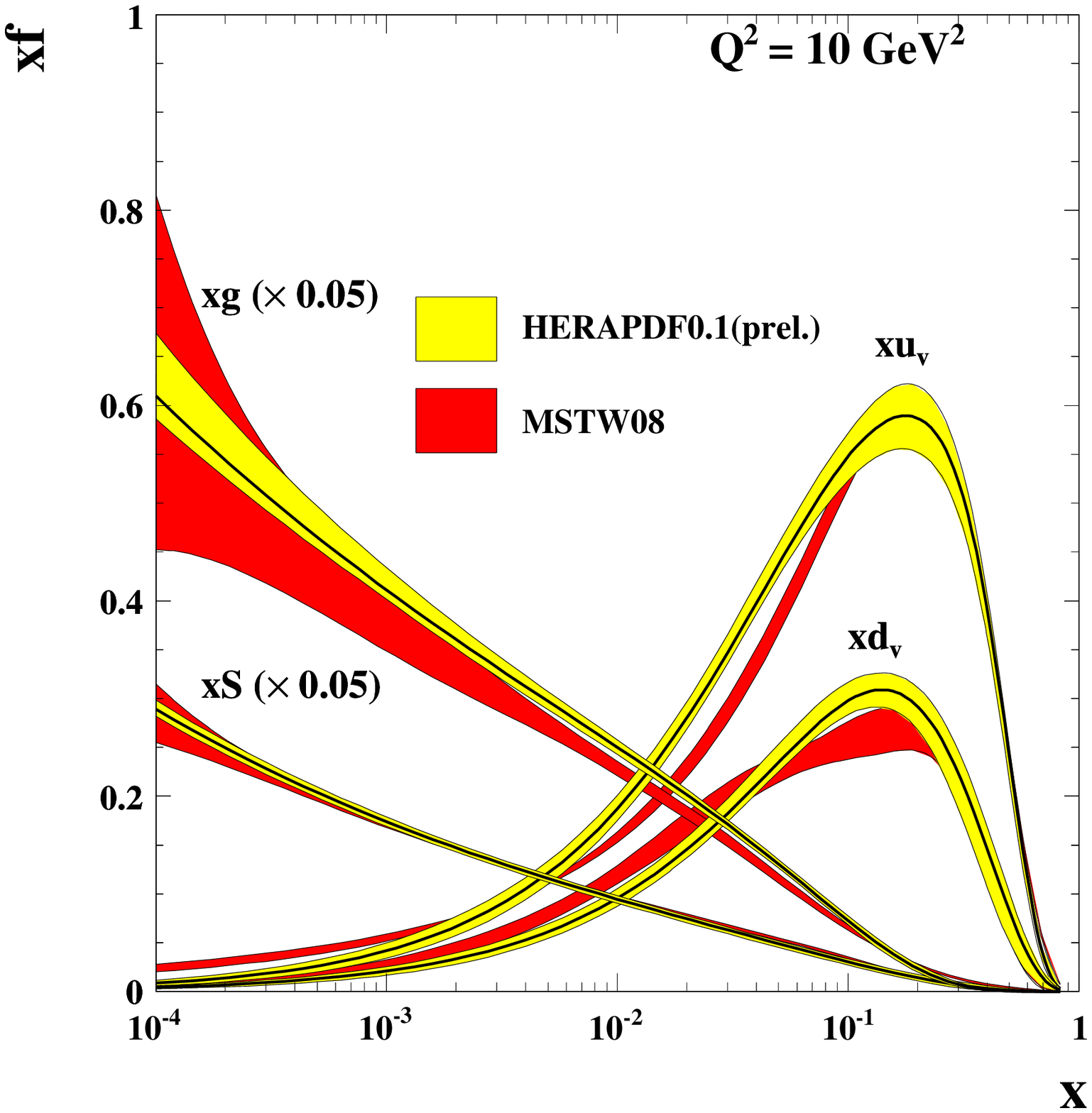,width=0.5\textwidth,height=6.5cm}
}
\caption {Left: HERAPDF0.1 
at $Q^2=10$GeV$^2$ compared to the CTEQ6.1 PDFs. Right: HERAPDF0.1 at $Q^2=10$GeV$^2$ 
compared to the preliminary MSTW08 PDFs. 
}
\label{fig:cteqmstw}
\end{figure}

In Fig.~\ref{fig:cteqmstw} we show the HERAPDF0.1 PDFs compared to the
CTEQ6.1 PDFs, which also use a zero-mass variable flavour number scheme, 
and to the preliminary MSTW08 PDFs~\cite{watt}, which use a massive variable flavour 
number scheme. The precision of the HERAPDF0.1 for the low-$x$ 
sea and gluon is impressive.

\section{Summary}
\label{sec:conc}
Now that high-$Q^2$ HERA data on NC and CC
 $e^+p$ and $e^-p$ inclusive double 
differential cross-sections are available, PDF fits can be made to HERA 
data alone, since the HERA high $Q^2$ cross-section 
data can be used to determine the valence distributions and HERA low $Q^2$ 
cross-section data can be used to determine the Sea and gluon distributions. 
The combined HERA-I data set, of neutral and charged current 
inclusive cross-sections for $e^+p$ and $e^-p$ scattering, has been used as the
sole input for a NLO QCD PDF fit in the DGLAP formalism. 
The consistent treatment of systematic uncertainties in the joint data set 
ensures that experimental uncertainties on the PDFs can be calculated
without 
need for an increased $\chi^2$ tolerance. This results in PDFs with greatly
reduced experimental uncertainties compared to the separate analyses of 
the ZEUS and H1 experiments. Model uncertainties, including those 
arising from parametrization dependence, have also been carefully considered. 
The resulting HERAPDFs  
have impressive precision compared to the global fits.


\begin{footnotesize}



%

\end{footnotesize}


\end{document}